\documentclass[article,twocolumn]{revtex4-1}
\usepackage{graphicx}
\usepackage{dcolumn}
\usepackage{bm}
\usepackage{amsmath}
\let\oldAA\AA
\renewcommand{\AA}{\text{\normalfont\oldAA}}

\usepackage{amsmath}
\usepackage{subfigure}
\usepackage{hyperref}
\usepackage[normalem]{ulem}
\hypersetup{
    colorlinks,
    citecolor=blue,
    filecolor=black,
    linkcolor=blue,
    urlcolor=blue
}
\usepackage{xcolor}

\usepackage{epstopdf}
 \usepackage{relsize}
\usepackage{ragged2e}

\newcommand*{\rom}[1]{\expandafter\@slowromancap\romannumeral #1@}
\title{Report}

\setcitestyle{square}

\begin{document}


\title{Enhanced thermoelectric performance  actuated by inelastic processes in the channel region}

\author{Aniket Singha}

\affiliation{%
Department of Electronics and Electrical Communication Engineering,\\
Indian Institute of Technology Kharagpur, Kharagpur-721302, India\\
}%






\begin{abstract}
  I  propose a design strategy  to enhance the performance of heat engine via absorption of thermal energy from the channel region.  The absorption of thermal energy  can  be actuated by inelastic processes and may be accomplished by an energy restrictive flow of electrons into the channel. The proposed design strategy employs dual energy filters to inject and extract electrons through the contact to channel interface. The first filter injects a compressed stream of electrons from the hot contact, at an effective temperature much lower than the channel temperature. The compressed stream of injected electrons, then, absorb energy via inelastic scattering inside the channel and  are finally extracted via a second filter at the cold contact interface. Rigorous mathematical derivations demonstrate that an optimized performance for the proposed design strategy demands the implementation of a box-car transmission function at the electron injecting terminal and a unit-step transmission function at the  electron extracting terminal. Numerical simulation show that in the proposed design strategy, the  heat engine performance, under optimal conditions, can surpass the ballistic limit when the required output power is low compared to the quantum bound. The proposed concept can be used to construct high efficiency thermoelectric generators in situations where the source of usable heat energy is limited, but insulated from environmental dissipation.
\end{abstract}
\maketitle
\section{Introduction}
The efficiency of any heat engine is limited by the Carnot efficiency defined as:
\begin{equation}
\eta_C=1-T_{C}/T_H,
\end{equation}
 where $T_{H}$ and $T_C$ denote the temperatures of the hot and cold contacts respectively. Typically, the Carnot efficiency is achieved at zero power output and vanishing lattice thermal conductivity ($\kappa_{ph}$). Modern thermoelectrc engineering aims at enhancing the  \emph{figure of merit} ($zT$) defined as:
\begin{equation}
zT=\frac{S^2\sigma}{\kappa}T,
\label{eq:zt}
\end{equation}
where $S$, $\sigma$ and $\kappa$ are the Seebeck coefficient, the electrical conductivity and the thermal conductivity of the material respectively, and $T$ is the average temperature between the hot and cold contacts.  In the linear response regime, the operation efficiency of a thermoelectric generator is closely related to the \emph{figure of merit} $zT$. In  attempts towards enhancing $zT$, the two approaches commonly followed are $(i)$ limiting the thermal conductivity $\kappa$  and $(ii)$ enhancing the power factor ($S^2\sigma$). The thermal conductivity $\kappa$ captures the heat flow due to lattice thermal conductivity ($\kappa_{ph}$) as well as  due to electronic thermal conductivity ($\kappa_{el}$). To enhance the efficiency, a lot of engineering effort has been dedicated towards reducing  $\kappa_{ph}$. Such efforts include  nanostructuring, nano-inclusions,  embedded interfaces and heterostructures \cite{supressk1,supressk2,phonon1,phonon2,phonon3,phonon4,phonon5,chen3}. An engineering of electronic density of states, on the other hand, is an independent and alternative route that also aims at enhancing the power factor ($S^2\sigma$), while simultaneously reducing $\kappa_{el}$ as far as possible \cite{sothmann1,sothmann2,staircase_qdot,aniket1,aniket,aniket2,aniket3,rev_sofomahan,Humphrey,Humphrey2,yamamoto2,yamamoto1}. \\
\indent In thermoelectric generators, the overall generation efficiency is generally limited by the lattice thermal conductivity ($\kappa_{ph}$). The heat energy flowing from the hot  contact via lattice thermal conductivity,  either dissipates to the environment from the channel region or flows towards the cold contact and can't be re-extracted back to the hot contact. This results in a deterioration of the overall efficiency. The overall Seebeck coefficient, hence, the power factor and efficiency, could be improved, if by clever engineering of electronic density of states, the heat flowing into channel region via lattice heat conductivity, could be absorbed by the electrons. In other words, the overall Seebeck coefficient can be improved, while still limiting the electronic heat conductivity, via inelastic processes in the channel region. A few proposals regarding the enhancement of the Seebeck coefficient or power factor  via inelastic scattering in the channel  region \cite{kim1,kim2,neo} has  been put forward  in recent years. However, we are yet to see a powerful, but   general and compact design strategy to enhance the heat engine performance via inelastic processes. In this paper, I construct a general, but compact  design strategy for maximum enhancement in thermoelectric performance via absorption of thermal energy from  the channel. The optimal design features of the proposal were derived via rigorous mathematical calculations. Along with optimizing the design features, I also present numerical simulation results to assess the enhancement in thermoelectric performance via my proposed strategy. \\
\indent The absorption of waste heat from the channel region can be facilitated by restricting the energy resolved flow of electrons from the hot contact, which in other words is known as energy filtering \cite{kim1,kim2,aniket,aniket3}. Energy filtering of incoming electrons in the channel region can be accomplished via  control of the transmission coefficient at the hot contact to channel junction. This  results in the injection of electrons with an effective temperature  much lower than the channel temperature. It should be noted that absorption of heat from the channel doesnot impact the heat flowing  from the hot contact via lattice heat conductivity, since the direction of heat flow is from the channel towards the cold contact and not vice-versa. In this context, it might be stated that the channel temperature is intermediate between the hot and cold contacts. It is worth mentioning that, unlike the given Refs. \cite{kim1,kim2,neo}, where much of the enhancement in generated power stems from an enhanced  electrical conductivity $\sigma(\epsilon)$ at higher energy, the proposed concept in this paper is independent of such constraints. Rather, in this case, I aim towards finding out the optimal bounded transmission functions   for maximizing the efficiency at a  generated power. While doing so, I show that the efficiency of the heat engine with respect to the electronic heat flow, can surpass the ballistic limit at finite power output, provided that the transmission functions are optimized to achieve the best performance. \\
\indent This paper is organized as follows. In Sec.~\ref{model}, I highlight the model used to validate the impact of the proposed design strategy. Sec \ref{calculation} consists of detailed calculation to optimize the design features of the model, proposed in Sec. \ref{model}, to achieve maximum enhancement in generated power. In Sec. \ref{results}, I demonstrate result of numerical simulations for the proposed design strategy, considering the model discussed in  Sec.~\ref{model}. I, finally, conclude this paper with a general discussion in Sec.~\ref{conclusion}.
 
\section{Model.}\label{model}
\begin{figure}[!tb]
{\includegraphics[scale=.25]{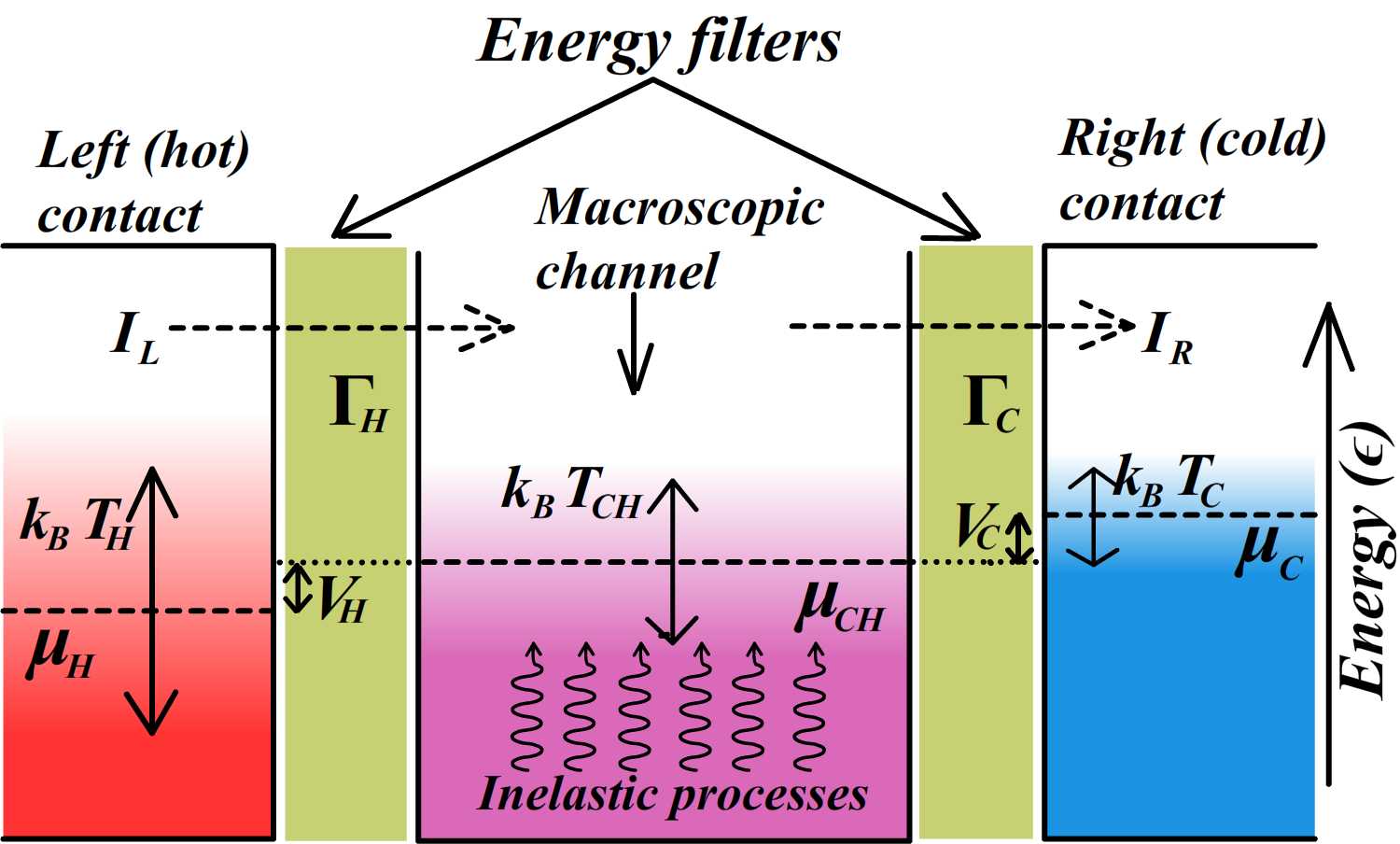}}
\caption{Schematic diagram illustrating the model employed to validate the proposed design strategy. The model consists of a left (hot) contact at temperature $T_H$ and  a right (cold) contact at temperature $T_C$ connected by a middle channel region at temperature $T_{CH}$. Two electron filters with transmission function $\Gamma_H$ and $\Gamma_C$ are connected at the hot contact to channel and cold contact to channel interface.  The filters dictate the energy resolved electronic current flowing through the interfaces. Inelastic processes within the channel may be used to facilitate heat absorption by the incoming electrons, provided that the effective temperature of the injected electrons is lower compared to the channel.}
\label{fig:schem}
\end{figure}

To validate the proposed concept, I employ a toy model as shown in Fig.~\ref{fig:schem}. The model  consists  of a left hot contact at temperature $T_H$ and a right cold contact at temperature $T_C$. The contacts are connected to the middle channel region labeled `CH' via electronic filters with energy dependent transmission function $\Gamma_H$ and $\Gamma_C$. The channel, in this case, is assumed to be a macroscopic electronic bath with well defined quasi-Fermi energy $\mu_{CH}$. Such an assumption is valid  when the resistance of the channel is much less compared to that of the energy filters $\Gamma_H$ and $\Gamma_C$. This basically means that the electronic flow between the hot (left) and cold (right) contacts is practically dictated by the two filters at the channel-to-contact junctions.  For ease of analysis, I also assume that there is no spatial variation of the channel temperature, labeled as $T_{CH}$, which  is valid when the thermal conductivity of the filters are much lower compared to that of the channel \cite{kim1,kim2}.  In addition to dictating the electronic flow, the left filter ($\Gamma_H$) also governs the electronic heat current that flows from the hot contact. By a suitable choice of the energy resolved transmission coefficient $\Gamma_H$ of the filter at the hot contact to channel junction, electrons can be injected at an effective temperature that is lower compared to the channel. This in turn facilitates an absorption of heat energy from the channel via inelastic processes. \\
\indent In the linear response regime,  the power factor $S^2G$ and  the figure of merit $zT$ can be used to assess the performance of a thermoelectric generator. However, these parameters are not of much ultility when the corresponding generator is operating in the non-linear regime \cite{nakpathomkun}. In addition, these two parameters hardly give any information about the regimes of operation other than the point of maximum power generation, and hence, cannot facilitate a clear understanding of the physics of heat flow \cite{Humphrey,jordan1,jordan2,sothmann,bm,whitney,whitney2}.     Hence, an analysis of power generation at a given  efficiency and operating point \cite{whitney,whitney2,sothmann,bm,agarwal,BD,nakpathomkun,jordan2,bm,zimb,leij} is essential for a detailed analysis of the advantage gained via the proposed concept. In this paper, I hence, carry out an analysis of generated power at a given efficiency.\\
\indent Here, a direct calculation of maximum  power at a given efficiency  is performed by optimizing the filter transmission functions $\Gamma_H$ and $\Gamma_C$. The voltage drop across an external load, where the power dissipation takes place, is emulated by a potential bias across the generator \cite{nakpathomkun}. The generated power $(P)$ and the efficiency ($\eta$) for a  a given temperature  and voltage bias across two contacts can be calculated as:
\begin{equation}
P=I \times V
\label{eq:pow}
\end{equation} 
\begin{equation}
\eta=\frac{P}{I_Q},
\label{eq:eff}
\end{equation}
where $I$ is the electronic charge current, $I_Q$ is the total heat current at the hot contact and the applied bias $V$ emulates the potential drop across the external load.   Since lattice thermal conductivity remains almost unchanged regardless of the electronic heat flow, it is logical to simplify our calculations by neglecting the degradation in efficiency due to phonon heat conductivity. In addition, reduction of lattice heat conductivity via engineering of nano-heterostructures is a different and independent path towards improving the overall efficiency of waste heat harvesting. So, the efficiency  that I use to gauge the effectiveness of the proposed concept   is  defined  by the equation:
\begin{equation}
\eta=\frac{P}{I_{Qe}},
\label{eq:ef}
\end{equation}
where $P$ is  defined in \eqref{eq:pow} and $I_{Qe}$ is the electronic heat current at the hot contact.   As stated earlier, I assume that the resistance of the  channel `CH' is negligible compared to the energy filters. Hence, the electronic current between the contacts is effectively limited by the energy filters. This also means that the total voltage drop across the channel is negligible compared to that across the filters. I denote the respective electrochemical potentials of the hot contact,  cold contact and  the macroscopic channel by  $\mu_H$, $\mu_C$ and $\mu_{CH}$ respectively. Without loss of generality, I assume that $\mu_{CH}=0$. Assuming a voltage drop of $V_H$ and $V_C$  entirely across the left and right energy filters respectively, the quasi-Fermi levels of the hot and the cold contacts are  given by $\mu_H=-eV_H$ and $\mu_C=eV_C$ (assuming no spacial variation in $\mu_{CH}$). The total voltage drop across the generator is hence given by $V=V_H+V_C$. Under the condition of ballistic electronic transport  through the two filters, the electronic charge currents through the left and right filters  can be calculated from Landauer's scattering theory via the following equations: 
\begin{eqnarray}
I_L=\frac{e}{h} \int \Gamma _H(\epsilon) \{f_H(\epsilon)-f_{CH}(\epsilon)\} d\epsilon \nonumber \\
I_R=\frac{e}{h} \int \Gamma _C(\epsilon) \{f_{CH}(\epsilon)-f_C(\epsilon)\} d\epsilon
\label{eq:curr}
\end{eqnarray}
In steady state, $I_L=I_R=I$.  The net irreversible heat flow within the system, on the other hand, occurs from the hot contact to the channel and subsequently towards the cold contact. Hence, the net irreversible heat flow can be accounted for by considering only the heat lost from the hot contact (since the heat lost from the channel to the cold contact ultimately comes from the hot contact via lattice heat conduction or electronic heat conduction). Hence, for the purpose of my calculations, the net irreversible  heat current  flow in the system  is defined as the net heat flow between the left (hot) contact to channel  and depends only on  the transmission function $\Gamma_H$.  As stated before, for the purpose of calculations, I only consider the heat flow due to the electronic conduction. The net irreversible heat flow from the left (hot) contact can, hence, be defined by the equation:
\begin{equation}
I_{Qe}=\frac{1}{h} \int (\epsilon-\mu_H)\Gamma _H(\epsilon) \{f_H(\epsilon)-f_{CH}(\epsilon)\} d\epsilon
\label{eq:heat}
\end{equation}
In the above equations, $f_H(\epsilon)$, $f_C(\epsilon)$ and $f_{CH}(\epsilon)$ denote the quasi Fermi electronic distribution at the hot contact, the cold contact and the macroscopic electronic channel `CH' respectively.
\[
f_i=\left[1+exp\left(\frac{\epsilon-\mu_i}{k_B T_i}\right)\right]^{-1}
\] 
\indent The set of equations are written  under the assumption that the electrons injected from the left contact completely equilibriate with the channel temperature $T_{CH}$ via inelastic scattering while transport through the channel. It should be noted that the absorption of heat from the channel hardly impacts the electronic or lattice heat flow from the left contact. This is because both electronic and lattice heat flows from the hot contact towards the channel and subsequently cold contact, and not in the reverse direction. Hence,  absorption of heat via inelastic processes within the channel simply reduces the lattice heat flux into the cold contact. Throughout the discussion, I will assume  n-type channel, though the discussion is  valid for p-type channel as well, with a slight modification of the equations used in the following sections.
\section{Calculation and optimization of $\Gamma_H$ and $\Gamma_C$.}\label{calculation}
In this section, I present analytical calculations for  $\Gamma_H$ and $\Gamma_C$ to maximize the generated power at a given efficiency. Since, the transmission function of any electronic filters can be complicated, I fix the upper bounds on the functions $\Gamma_H$ and $ \Gamma_C$ at a given energy $\epsilon$, assuming that by clever engineering the filters can be manipulated to acquire the desired transmission characteristics. The  upperbound to the transmission functions are set such that:
\begin{equation}
0\leq \Gamma_H(\epsilon)\leq 1;  ~0 \leq \Gamma_C(\epsilon) \leq 1,
\end{equation}
for any energy $\epsilon$ in the range of electronic transport. My intention here is to derive the mathematical form  of the  optimal transmission function for  the filters $\Gamma_H$ and $\Gamma_C$. It should be noted in this regard that the derived concepts and results still hold if one multiplies the optimized transmission function by a constant.
\\
\indent For  analysis of the problem,  I define a variable, 
\begin{equation}
I_C=I_L-I_R
\end{equation}
In steady state, $I_C=0$.  I  assume that the transmission function of each of the filters $\Gamma _H(\epsilon)$ and $\Gamma _C(\epsilon)$ can take any value from $0$ to a maximum value of $1$ and proceed towards calculating the maximum power at a given efficiency. My approach to the problem entails finding out the maximum value of the power $P$ at a given value of the electronic heat current at the left (hot) contact $I_{Qe}$. The generated power $P$ and the electronic heat current $I_{Qe}$ are functions of $V_H$, $V_C$, $\Gamma_H$ and $\Gamma_C$. Hence, a small variation in $P$ and $I_{Qe}$ due to small variation in this parameter space can be written as: 
\indent 
\begin{widetext}
\begin{equation}
\delta P= \sum_i\delta \Gamma _i(\epsilon)\frac{\partial P}{\partial \Gamma_i(\epsilon)}\Big|_{V_H, V_C}+\delta V_H\frac{\partial P}{\partial V_H}\Big|_{V_C, \Gamma}+\delta V_C\frac{\partial P}{\partial V_C}\Big|_{V_H, \Gamma},
\label{eq:power_differential}
\end{equation}

\begin{equation}
\delta I_{Qe}= \sum_i \delta \Gamma _i(\epsilon)\frac{\partial I_{Qe}}{\partial \Gamma_i(\epsilon)}\Big|_{V_H, V_C}+\delta V_H\frac{\partial I_{Qe}}{\partial V_H}\Big|_{V_C, \Gamma}+\delta V_C\frac{\partial I_{Qe}}{\partial V_C}\Big|_{V_H, \Gamma},
\label{eq:heat_differential}
\end{equation}
\end{widetext}
where  the symbol ~~`$|_{x}$'~~ implies that the partial differentiation is carried out at constant values of $x$ and $i \in (H,~C)$, $\Gamma=(\Gamma_H, \Gamma_C)$. Since $I_{Qe}$, in \ref{eq:heat} doesn't depend explicitly on $V_C$ and $\Gamma_C$, I put $\frac{\partial I_{Qe}}{\partial V_C}\Big|_{V_H, \Gamma}=\frac{\partial I_{Qe}}{\partial \Gamma_C}\Big|_{V_H, V_C}=0$. Hence,

\begin{equation}
\delta I_{Qe}= \delta \Gamma _H(\epsilon)\frac{\partial I_{Qe}}{\partial \Gamma_H(\epsilon)}\Big|_{V_H, V_C}+\delta V_H\frac{\partial I_{Qe}}{\partial V_H}\Big|_{V_C, \Gamma}.
\label{eq:heat_differential_rewrite}
\end{equation}
Similarly,
\begin{eqnarray}
\delta I_C=\sum_i \delta \Gamma _i(\epsilon)\frac{\partial I_C}{\partial \Gamma_i(\epsilon)}\Big|_{V_H, V_C}+\delta V_H\frac{\partial I_C}{\partial V_H}\Big|_{V_C, \Gamma} \nonumber \\ +\delta V_C\frac{\partial I_C}{\partial V_C}\Big|_{V_H, \Gamma}~~~~~~~~
\label{eq:curr_differential_rewrite}
\end{eqnarray}
At first, I assume $\Gamma_H$ to be some arbitrary function and proceed towards finding out the optimal bounded function for $\Gamma_C$. While $\Gamma_C$ varies, I try to maximize the generated power for a given heat current. Hence, I  fix $I_{Qe}$ and vary the parameters $\Gamma _C(\epsilon)$, $V_H$ and $V_C$ so that the output power  varies along a line where $I_{Qe}$ is kept constant in the parameter ($\Gamma _C(\epsilon),~V_C,~V_H$) space. In this case, I can rewrite Eqs.  \eqref{eq:heat_differential_rewrite} and \eqref{eq:curr_differential_rewrite} as:
\begin{equation}
\delta I_{Qe}= \delta V_H\frac{\partial I_{Qe}}{\partial V_H}\Big|_{V_C, \Gamma}=0 
\label{eq:right_solve_1}
\end{equation} 
\begin{eqnarray}
\delta I_C= \delta \Gamma _C(\epsilon)\frac{\partial I_C}{\partial \Gamma_C(\epsilon)}\Big|_{V_H, V_C}+\delta V_H\frac{\partial I_C}{\partial V_H}\Big|_{V_C, \Gamma} \nonumber \\ +~~\delta V_C\frac{\partial I_C}{\partial V_C}\Big|_{V_H, \Gamma}=0,~~~~~
\label{eq:right_solve_2}
\end{eqnarray}
Solving equations \eqref{eq:right_solve_1} and \eqref{eq:right_solve_2} I get the values of $\delta V_H$ and $\delta V_C$ as:
\begin{equation}
\delta V_C=-\delta \Gamma _C(\epsilon) \frac{\frac{\partial I_C}{\partial \Gamma_C(\epsilon)}\Big|_{V_H, V_C}}{\frac{\partial I_C}{\partial V_C}\Big|_{V_H, \Gamma}},~~\delta V_H=0.
\label{eq:delta_1}
\end{equation}
Substituting  the values of $\delta V_H$ and $\delta V_C$ in \eqref{eq:power_differential} from \eqref{eq:delta_1} I get,
\begin{equation}
\delta P=\delta \Gamma _C(\epsilon)\frac{\partial P}{\partial \Gamma_C(\epsilon)}\Big|_{V_H, V_C}-\delta \Gamma _C(\epsilon) \frac{\frac{\partial I_C}{\partial \Gamma_C(\epsilon)}\Big|_{V_H, V_C}}{\frac{\partial I_C}{\partial V_C}\Big|_{V_H, \Gamma}}\frac{\partial P}{\partial V_C}\Big|_{V_H, \Gamma},
\label{eq:delta_2}
\end{equation}
To proceed from the above equation, I substitute $\frac{\partial I_C}{\partial \Gamma_C(\epsilon)}\Big|_{V_H, V_C}$ in \eqref{eq:delta_2} by the formulas :
\begin{equation}
\frac{\partial I_C}{\partial \Gamma_C(\epsilon)}\Big|_{V_H, V_C}=-\frac{\partial I_R}{\partial \Gamma_C(\epsilon)}\Big|_{V_H, V_C}=-\frac{1}{V_C}\frac{\partial P}{\partial \Gamma_C(\epsilon)}\Big|_{V_H, V_C}
\end{equation}
\begin{equation}
\frac{\partial P}{\partial V_C}=I_R+V_C \frac{\partial I_R}{\partial V_C}=I_R-V_C\frac{\partial I_C}{\partial V_C} 
\end{equation}
I hence get:
\begin{align}
\delta P&=\left\{1+ \frac{1}{V_C}\frac{\frac{\partial P}{\partial V_C}\Big|_{V_H, V_C}}{\frac{\partial I_C}{\partial V_C}\Big|_{V_H, \Gamma}}\right\}\delta \Gamma _C(\epsilon)\frac{\partial P}{\partial \Gamma_C(\epsilon)}\Big|_{V_H, V_C} \nonumber \\
&=\left\{1+ \frac{I_R}{V_C {\frac{\partial I_C}{\partial V_C}}}-1\right\} \delta \Gamma _C(\epsilon)\frac{\partial P}{\partial \Gamma_C(\epsilon)}\Big|_{V_H, V_C}\nonumber \\
&=\frac{I_R}{V_C {\frac{\partial I_C}{\partial V_C}}}\delta \Gamma _C(\epsilon)\frac{\partial P}{\partial \Gamma_C(\epsilon)}\Big|_{V_H, V_C}
\end{align}
Now, $\frac{\partial I_C}{\partial V_C}=-\frac{\partial I_R}{\partial V_C}>0$ (from Eq. \ref{eq:curr}). Assuming an n-type channel and $\mu_{CH}=0$, it can be readily found out that $\frac{\partial P}{\partial \Gamma_C (\epsilon)}\geq 0$ for $\epsilon\geq \epsilon_C$, with $\epsilon_C=\mu_C \left(1-\frac{T_C}{T_{CH}}\right)^{-1}$. Hence, the generated power increases for a small increase in $\Gamma_C(\epsilon)$ by  $\delta \Gamma _C(\epsilon)$, when $\epsilon \geq \epsilon_C$. For $\epsilon<\epsilon_C$, the generated power decreases with a small increase in $\Gamma_C(\epsilon)$. Hence, to generate maximum power for a given value of the $I_{Qe}$, the transmission function $\Gamma_C$ should to kept to its maximum and minimum possible value for $\epsilon \geq \epsilon_C$ and $\epsilon < \epsilon_C$ respectively.
 Hence, in this particular case, where $0 \leq \Gamma_C(\epsilon) \leq 1$, the mathematical expression for $\Gamma_C$ to generate maximum power is:
\begin{equation}
\Gamma_C=\theta(\epsilon-\epsilon_C),
\label{eq:final_gamma_c}
\end{equation}
$\theta$ being the unit step function.\\
\indent Next, I proceed towards finding out the optimized transmission function for the left filter. We first note that the transmission function $\Gamma_C$ given by Eq. \eqref{eq:final_gamma_c} is independent for $\Gamma_H$ for any value of $\epsilon$.  Hence, I  assume that the transmission function for the right filter has already been fixed to its optimal value. If the transmission function $\Gamma_H$, at energy $\epsilon$, changes by a small amount  $\delta \Gamma_H (\epsilon)$, then for a fixed value of $I_{Qe}$, Eq. \eqref{eq:heat_differential} becomes:

\begin{equation}
\delta I_{Qe}= \delta \Gamma _H(\epsilon)\frac{\partial I_{Qe}}{\partial \Gamma_H(\epsilon)}\Big|_{V_H, V_C}+\delta V_H\frac{\partial I_{Qe}}{\partial V_H}\Big|_{V_C, \Gamma}=0,
\label{eq:left_solve_1}
\end{equation}
\begin{eqnarray}
\delta I_C= \delta \Gamma _H(\epsilon)\frac{\partial I_C}{\partial \Gamma_H(\epsilon)}\Big|_{V_H, V_C}+\delta V_H\frac{\partial I_C}{\partial V_H}\Big|_{V_C, \Gamma}\nonumber \\+\delta V_C\frac{\partial I_C}{\partial V_C}\Big|_{V_H, \Gamma} =0
\label{eq:left_solve_2}
\end{eqnarray}

Solving \eqref{eq:left_solve_1} and \eqref{eq:left_solve_2}, I get 
\begin{equation}
\delta V_H=-\frac{\frac{\partial I_{Qe}}{\partial \Gamma_H(\epsilon)}\Big|_{V_H,~V_C}}{\frac{\partial I_{Qe}}{\partial V_H}\Big|_{\Gamma, V_C}}\delta \Gamma _H (\epsilon) \\
\label{eq:left_solve_3}
\end{equation}
\begin{equation}
\delta V_C=\left(\frac{\frac{\partial I_{Qe} }{\partial \Gamma _H(\epsilon)}\Big|_{V_H,V_C} \frac{\partial I_C}{\partial V_H}\Big|_{V_C,\Gamma_H}}{\frac{\partial I_{Qe}}{\partial V_H}\Big|_{V_C,\Gamma_H} \frac{\partial I_C}{\partial V_C}\Big|_{V_H,\Gamma_H}}-\frac{\frac{\partial I_C}{\partial \Gamma _H}\Big|_{V_H,V_C}}{\frac{\partial I_C}{\partial V_C}\Big|_{V_H,\Gamma_H}}\right)\delta \Gamma_H(\epsilon)
\label{eq:left_solve_4}
\end{equation}
\indent Note that $\delta V_H$ in Eq. \eqref{eq:left_solve_3} depends only on the parameters $V_H$ and $\Gamma_H(\epsilon)$. This basically means that the optimization of $\Gamma_H(\epsilon)$, and hence $V_H$, at a given value of $I_{Qe}$ are independent of $V_C$. 
I hence independently optimize the power $P_H$ given by $P_H=I_L \times V_H$. The change in power $P_H$ due to a small change in $\Gamma_H(\epsilon)$ and $V_H$ is given by:
\begin{equation}
\delta P_H=\delta \Gamma_H(\epsilon)\frac{\partial P_H}{\partial \Gamma_H(\epsilon)}\Big|_{V_H, V_C}+\delta V_H\frac{\partial P_H}{\partial V_H}\Big|_{V_C, \Gamma}
\label{eq:delta_P_H}
\end{equation}
Substituting the value of $\delta V_H$ in the above equation from \eqref{eq:left_solve_3} and replacing $\frac{\partial I_{Qe}}{\partial \Gamma_H(\epsilon)}$ using the formula,
\begin{equation}
\frac{\partial I_{Qe}}{\partial \Gamma_H(\epsilon)}\Big|_{V_H, V_C}=\frac{\epsilon-\mu_H}{eV_H}\frac{\partial P_H}{\partial \Gamma_H(\epsilon)}\Big|_{V_H, V_C},
\end{equation}
Eq. \eqref{eq:delta_P_H} can be re-written as,
\begin{equation}
\delta P_H=\delta \Gamma_H(\epsilon)\frac{\partial P_H}{\partial \Gamma_H(\epsilon)}\Big|_{V_H, V_C}\left\{ 1-\frac{\epsilon-\mu_H}{eV_H}\frac{\frac{\partial P_H}{\partial V_H}\Big|_{V_C, \Gamma}}{\frac{\partial I_{Qe}}{\partial V_H}\Big|_{V_C, \Gamma}}\right\}
\label{eq:delta_P_H_2}
\end{equation}

In Eq. \eqref{eq:delta_P_H_2}, I observe that $\frac{\partial P_H}{\partial \Gamma_H(\epsilon)}>0$ for $\epsilon>\epsilon_H$, where,
\begin{equation}
 \epsilon_H=-\mu_H\left(\frac{T_H}{T_{CH}}-1\right)^{-1}.
 \label{eq:jye_h}
 \end{equation}
 
  In addition, $I_{Qe}$ is a decreasing function of the applied voltage $V_H$, and hence  $\frac{\partial I_{Qe}}{\partial V_H}\Big|_{V_C, \Gamma}<0$	 (from Eq. \ref{eq:heat}). $P_H$ on the other hand is a non-monotonic function of the applied voltage $V_H$. $P_H$ increases from zero when $V_H$ increases from $V_H=0$ and then reaches a point when $\frac{\partial P_H}{\partial V_H}\Big|_{V_C, \Gamma}=0$ and then again decreases making $\frac{\partial P_H}{\partial V_H}\Big|_{V_C, \Gamma}<0$. Eq. \eqref{eq:delta_P_H_2} demonstrates that for $\frac{\partial P_H}{\partial V_H}\Big|_{V_C, \Gamma}>0$, $\delta P_H$ is positive for any  $\epsilon>\epsilon_H$. The challenge for optimization of $\Gamma_H(\epsilon)$ sets in when finally $\frac{\partial P_H}{\partial V_H}\Big|_{V_C, \Gamma}$ becomes negative. In this case, an increase in the power $\delta P_H$ is positive only if
\begin{widetext}

\begin{gather}
\left\{ 1-\frac{\epsilon-\mu_H}{eV_H}\frac{\frac{\partial P_H}{\partial V_H}\Big|_{V_C, \Gamma}}{\frac{\partial I_{Qe}}{\partial V_H}\Big|_{V_C, \Gamma}}\right\}=\left\{1-\frac{\frac{\partial P_H}{\partial V_H}\Big|_{V_C, \Gamma}}{\frac{\partial I_{Qe}}{\partial V_H}\Big|_{V_C, \Gamma}}-\frac{\epsilon}{eV_H}\frac{\frac{\partial P_H}{\partial V_H}\Big|_{V_C, \Gamma}}{\frac{\partial I_{Qe}}{\partial V_H}\Big|_{V_C, \Gamma}}\right\}>0,\nonumber \\
\Rightarrow \epsilon <\epsilon_H'= eV_H\left\{\frac{\frac{\partial I_{Qe}}{\partial V_H}\Big|_{V_C, \Gamma}}{\frac{\partial P_H}{\partial V_H}\Big|_{V_C, \Gamma}}-1\right\}
\label{eq:jye_h_d}
\end{gather}

\end{widetext}
Hence, from Eq.~\eqref{eq:jye_h} and \eqref{eq:jye_h_d},  we note that an increase in $\Gamma_H$ at energy $\epsilon$ by an infinitesimal quantity $\delta \Gamma_H(\epsilon)$, results in an enhancement in power generation only when
\begin{equation}
\epsilon_H\leq \epsilon \leq \epsilon_H' \nonumber
\end{equation}
Hence, the maximum generated power for a given heat current is achieved when the transmission function is set to its maximum possible value in the  the  energy range between $\epsilon_H$ and $\epsilon_H'$. For other energy range in the transport window, the transmission function should be set to the minimum possible value. In this particular case, where $0 \leq \Gamma_H(\epsilon) \leq 1$, the transmission function in the energy range between $\epsilon_H$ and $\epsilon_H'$ should be set to 1. For other energy range $\Gamma_H$ should be set to 0. Hence , the optimal function for $\Gamma_H$ can be given by:
\begin{equation}
\Gamma_H=\theta(\epsilon-\epsilon_H)-\theta(\epsilon-\epsilon_H'),
\end{equation}
where $\theta$ is the unit step function. 
\begin{figure}[!htb]
\includegraphics[scale=1.6]{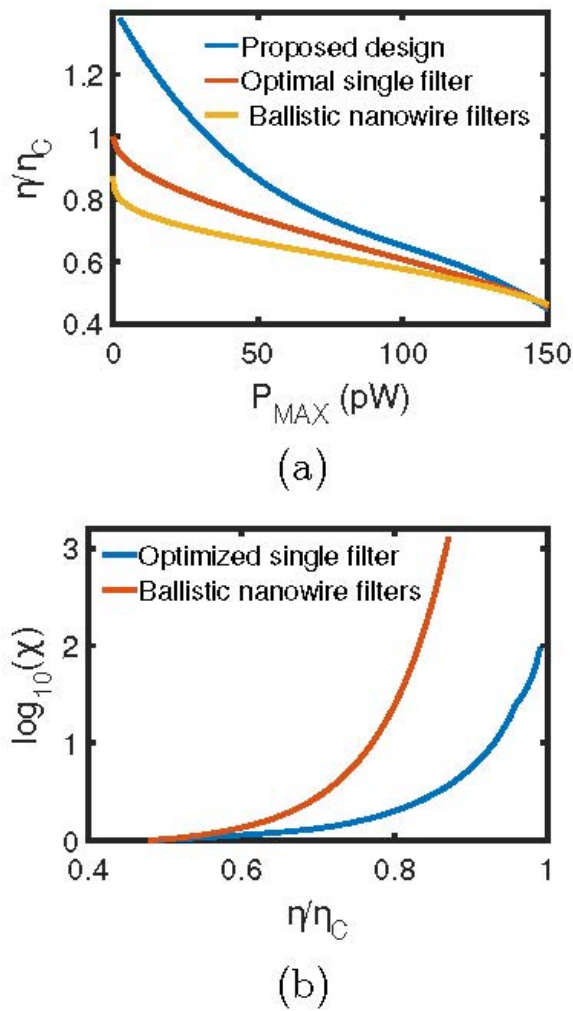}
\caption{(a) Maximum power ($P_{MAX}(\eta)$) vs efficiency ($\frac{\eta}{\eta_C}$) for our proposed design strategy, optimized single filter design proposed in Refs. \cite{whitney,whitney2} and ballistic nanowire filters. (b) Logarithm of the enhancement factor ($\chi$) with our proposed design strategy compared to a a single  optimized  filter design (boxcar transmission) proposed in Refs. \cite{whitney,whitney2} and ballistic nanowire filters. The efficiency plotted in the above figures are normalized with respect to the Carnot efficiency. For comparison of the optimal single filter design with the proposed design employing dual filter, the maximum value transmission function at any energy $\epsilon$ for the optimized single filter is limited to $0.5$. }
\label{fig:result}
\end{figure}
\section{Results}\label{results}
In this section, I demonstrate numerical simulation results for the case when $T_H=360K$, $T_{CH}=330K$ and $T_C=300K$. In this case, the values of the quantities $\epsilon_C$,~$\epsilon_H$ and $\epsilon_H'$ were numerically computed, followed by a numerical computation of the maximum generated power $P_{MAX}$ for a given efficiency $\eta$. For comparison, I normalize the efficiency with respect to the Carnot efficiency. To assess the performance enhancement via the proposed design strategy, I also demonstrate (in Fig.~\ref{fig:result}a) the results for (i) an optimal single filter design with boxcar transmission function (proposed in Refs. \cite{whitney,whitney2}) and (ii)  the filters $\Gamma_H$ and $\Gamma_C$ replaced by ballistic nanowires with transmission functions $\Gamma_H=\theta(\epsilon-\epsilon_H)$ and $\Gamma_C=\theta(\epsilon-\epsilon_C)$.  Fig.~\ref{fig:result}(a) demonstrates the maximum power generated at a given efficiency for these three cases.  We note that the advantage gained by the proposed design strategy is quite prominent in the high efficiency regime of operation, when the output power is low compared to the quantum bound \cite{whitney,whitney2}. In fact, when the desired output power is low compared to the quantum bound, the efficiency of operation, in the proposed design strategy, can be driven beyond the ballistic limit. However, such advantage gradually decreases as we  approach the point of maximum power generation. In particular, it should be noted that  all the three cases have identical magnitude of maximum generated power, in addition to identical efficiency at the point of maximum power. This is  expected, since $\epsilon_H'$ in our proposed model increase with the increase in generated power. Hence, the transmission function $\Gamma_H$ gradually approaches that of a ballistic nanowire when the generated power gradually approaches the quantum bound.\\
\indent To assess the enhancement in generated power at a given efficiency, I define a metric termed the \textit{advantage factor} ($\chi$), which is the ratio of the maximum generated power via our proposed design to the maximum  generated power via other designs.
\[ 
\chi(\eta)=\frac{P_{{MAX}}^{opt}(\eta)}{P_{MAX}^{other}(\eta)},
\]
where $P_{{MAX}}^{opt}(\eta)$ is the maximum power generated via the proposed design with optimal filters and $P_{MAX}^{other}(\eta)$ is the maximum generated power via other designs or strategies at efficiency $\eta$.  It should be noted that $\chi$ is a function of efficiency and is high in the regime of high efficiency.  Fig.~\ref{fig:result}(b) demonstrates the logarithm $\chi$ for two different cases, where the enhancement of generated power in our proposed design is compared with  (i) optimized single filter with boxcar transmission function as proposed in Refs.~\cite{whitney,whitney2}, and (ii) the same design shown in Fig.~\ref{fig:schem}, with the two filters replaced by ballistic nanowires, with $\Gamma_H=\theta(\epsilon-\epsilon_H)$ and $\Gamma_C=\theta(\epsilon-\epsilon_C)$. It should be noted that the enhancement of generated power near Carnot efficiency in our case is nearly $10^2$ times compared to the design proposed in Refs. \cite{whitney,whitney2}. The enhancement in generated power compared to ballistic nanowire filters is about $10^3$ times at $80\%$ of the Carnot efficiency, which is the maximum efficiency achieved via ballistic nanowire filters. It should, however, be noted that in this case the efficiency is defined with respect to electronic heat flow. In practice, however, the overall efficiency, defined with respect to the total heat flow due to electronic transport and lattice heat conductivity, should be much lower than the calculation result.
\section{Conclusion}\label{conclusion}
In this paper, I have  proposed a design strategy to enhance the performance of heat engines  via inelastic processes within the channel. In particular, I have proposed energy filters at the contact-to-channel interfaces for the said purpose. The energy filter at the hot contact to channel junction inject a stream of electrons with effective temperature much lower than the channel region. These electrons then absorb  thermal energy from the channel and flow towards the cold contact. Via rigorous mathematical derivations, it was shown that the heat engine reaches its peak performance (in case of electron-like conduction or n-type channel), when the filter at the hot-contact to channel junction  implements a box-car shaped transmission function, while that at the cold-contact to channel junction implements a unit-step shaped transmission function. Then, with the help of numerical simulation, I have  demonstrated that the principal utility of the proposed design strategy lies in the high efficiency regime of operation, where the efficiency of the engine, with respect to the electronic heat conduction, can be driven beyond the ballistic limit. Although, I have demonstrated the results for a particular case with   $T_H=360K$, $T_{CH}=330K$ and $T_C=300K$, the proposed concepts remain valid for other temperature range as well. In this paper, however, I have not considered an optimization of channel temperature for an optimal performance of the heat engine. It remains an interesting direction to explore the optimal channel temperature with respect to the contact temperatures for the proposed strategy.
The proposed concept in this paper can lead to the development of efficient heat engines, in cases where the source of usable thermal energy is limited.
\bibliography{References}

\providecommand{\noopsort}[1]{}\providecommand{\singleletter}[1]{#1}%
\begin{thebibliography}{34}%
\makeatletter
\providecommand \@ifxundefined [1]{%
 \@ifx{#1\undefined}
}%
\providecommand \@ifnum [1]{%
 \ifnum #1\expandafter \@firstoftwo
 \else \expandafter \@secondoftwo
 \fi
}%
\providecommand \@ifx [1]{%
 \ifx #1\expandafter \@firstoftwo
 \else \expandafter \@secondoftwo
 \fi
}%
\providecommand \natexlab [1]{#1}%
\providecommand \enquote  [1]{``#1''}%
\providecommand \bibnamefont  [1]{#1}%
\providecommand \bibfnamefont [1]{#1}%
\providecommand \citenamefont [1]{#1}%
\providecommand \href@noop [0]{\@secondoftwo}%
\providecommand \href [0]{\begingroup \@sanitize@url \@href}%
\providecommand \@href[1]{\@@startlink{#1}\@@href}%
\providecommand \@@href[1]{\endgroup#1\@@endlink}%
\providecommand \@sanitize@url [0]{\catcode `\\12\catcode `\$12\catcode
  `\&12\catcode `\#12\catcode `\^12\catcode `\_12\catcode `\%12\relax}%
\providecommand \@@startlink[1]{}%
\providecommand \@@endlink[0]{}%
\providecommand \url  [0]{\begingroup\@sanitize@url \@url }%
\providecommand \@url [1]{\endgroup\@href {#1}{\urlprefix }}%
\providecommand \urlprefix  [0]{URL }%
\providecommand \Eprint [0]{\href }%
\providecommand \doibase [0]{http://dx.doi.org/}%
\providecommand \selectlanguage [0]{\@gobble}%
\providecommand \bibinfo  [0]{\@secondoftwo}%
\providecommand \bibfield  [0]{\@secondoftwo}%
\providecommand \translation [1]{[#1]}%
\providecommand \BibitemOpen [0]{}%
\providecommand \bibitemStop [0]{}%
\providecommand \bibitemNoStop [0]{.\EOS\space}%
\providecommand \EOS [0]{\spacefactor3000\relax}%
\providecommand \BibitemShut  [1]{\csname bibitem#1\endcsname}%
\let\auto@bib@innerbib\@empty
\bibitem [{\citenamefont {Mahan}\ and\ \citenamefont
  {Sofo}(1996)}]{rev_sofomahan}%
  \BibitemOpen
  \bibfield  {author} {\bibinfo {author} {\bibfnamefont {G.~D.}\ \bibnamefont
  {Mahan}}\ and\ \bibinfo {author} {\bibfnamefont {J.~O.}\ \bibnamefont
  {Sofo}},\ }\href@noop {} {\ \textbf {\bibinfo {volume} {93}},\ \bibinfo
  {pages} {7436} (\bibinfo {year} {1996})}\BibitemShut {NoStop}%
\bibitem [{\citenamefont {{Whitney}}(2014)}]{whitney}%
  \BibitemOpen
  \bibfield  {author} {\bibinfo {author} {\bibfnamefont {R.~S.}\ \bibnamefont
  {{Whitney}}},\ }\href@noop {} {\bibfield  {journal} {\bibinfo  {journal}
  {Physical Review Letters}\ }\textbf {\bibinfo {volume} {112}},\ \bibinfo
  {eid} {130601} (\bibinfo {year} {2014})}\BibitemShut {NoStop}%
\bibitem [{\citenamefont {Whitney}(2015)}]{whitney2}%
  \BibitemOpen
  \bibfield  {author} {\bibinfo {author} {\bibfnamefont {R.~S.}\ \bibnamefont
  {Whitney}},\ }\href {\doibase 10.1103/PhysRevB.91.115425} {\bibfield
  {journal} {\bibinfo  {journal} {Phys. Rev. B}\ }\textbf {\bibinfo {volume}
  {91}},\ \bibinfo {pages} {115425} (\bibinfo {year} {2015})}\BibitemShut
  {NoStop}%
\bibitem [{\citenamefont {Androulakis}\ \emph {et~al.}(2006)\citenamefont
  {Androulakis}, \citenamefont {Hsu}, \citenamefont {Pcionek}, \citenamefont
  {Kong}, \citenamefont {Uher}, \citenamefont {D'Angelo}, \citenamefont
  {Downey}, \citenamefont {Hogan},\ and\ \citenamefont
  {Kanatzidis}}]{supressk1}%
  \BibitemOpen
  \bibfield  {author} {\bibinfo {author} {\bibfnamefont {J.}~\bibnamefont
  {Androulakis}}, \bibinfo {author} {\bibfnamefont {K.}~\bibnamefont {Hsu}},
  \bibinfo {author} {\bibfnamefont {R.}~\bibnamefont {Pcionek}}, \bibinfo
  {author} {\bibfnamefont {H.}~\bibnamefont {Kong}}, \bibinfo {author}
  {\bibfnamefont {C.}~\bibnamefont {Uher}}, \bibinfo {author} {\bibfnamefont
  {J.}~\bibnamefont {D'Angelo}}, \bibinfo {author} {\bibfnamefont
  {A.}~\bibnamefont {Downey}}, \bibinfo {author} {\bibfnamefont
  {T.}~\bibnamefont {Hogan}}, \ and\ \bibinfo {author} {\bibfnamefont
  {M.}~\bibnamefont {Kanatzidis}},\ }\href
  {http://dx.doi.org/10.1002/adma.200502770} {\bibfield  {journal} {\bibinfo
  {journal} {Advanced Materials}\ }\textbf {\bibinfo {volume} {18}},\ \bibinfo
  {pages} {1170} (\bibinfo {year} {2006})}\BibitemShut {NoStop}%
\bibitem [{\citenamefont {Hsu}\ \emph {et~al.}(2004)\citenamefont {Hsu},
  \citenamefont {Loo}, \citenamefont {Guo}, \citenamefont {Chen}, \citenamefont
  {Dyck}, \citenamefont {Uher}, \citenamefont {Hogan}, \citenamefont
  {Polychroniadis},\ and\ \citenamefont {Kanatzidis}}]{supressk2}%
  \BibitemOpen
  \bibfield  {author} {\bibinfo {author} {\bibfnamefont {K.~F.}\ \bibnamefont
  {Hsu}}, \bibinfo {author} {\bibfnamefont {S.}~\bibnamefont {Loo}}, \bibinfo
  {author} {\bibfnamefont {F.}~\bibnamefont {Guo}}, \bibinfo {author}
  {\bibfnamefont {W.}~\bibnamefont {Chen}}, \bibinfo {author} {\bibfnamefont
  {J.~S.}\ \bibnamefont {Dyck}}, \bibinfo {author} {\bibfnamefont
  {C.}~\bibnamefont {Uher}}, \bibinfo {author} {\bibfnamefont {T.}~\bibnamefont
  {Hogan}}, \bibinfo {author} {\bibfnamefont {E.~K.}\ \bibnamefont
  {Polychroniadis}}, \ and\ \bibinfo {author} {\bibfnamefont {M.~G.}\
  \bibnamefont {Kanatzidis}},\ }\href
  {http://science.sciencemag.org/content/303/5659/818} {\bibfield  {journal}
  {\bibinfo  {journal} {Science}\ }\textbf {\bibinfo {volume} {303}},\ \bibinfo
  {pages} {818} (\bibinfo {year} {2004})}\BibitemShut {NoStop}%
\bibitem [{\citenamefont {Mingo}\ and\ \citenamefont {Broido}(2004)}]{phonon1}%
  \BibitemOpen
  \bibfield  {author} {\bibinfo {author} {\bibfnamefont {N.}~\bibnamefont
  {Mingo}}\ and\ \bibinfo {author} {\bibfnamefont {D.~A.}\ \bibnamefont
  {Broido}},\ }\href {http://link.aps.org/doi/10.1103/PhysRevLett.93.246106}
  {\bibfield  {journal} {\bibinfo  {journal} {Phys. Rev. Lett.}\ }\textbf
  {\bibinfo {volume} {93}},\ \bibinfo {pages} {246106} (\bibinfo {year}
  {2004})}\BibitemShut {NoStop}%
\bibitem [{\citenamefont {Mingo}(2004)}]{phonon2}%
  \BibitemOpen
  \bibfield  {author} {\bibinfo {author} {\bibfnamefont {N.}~\bibnamefont
  {Mingo}},\ }\href
  {http://scitation.aip.org/content/aip/journal/apl/84/14/10.1063/1.1695629}
  {\bibfield  {journal} {\bibinfo  {journal} {Applied Physics Letters}\
  }\textbf {\bibinfo {volume} {84}},\ \bibinfo {pages} {2652} (\bibinfo {year}
  {2004})}\BibitemShut {NoStop}%
\bibitem [{\citenamefont {Zhou}\ \emph {et~al.}(2007)\citenamefont {Zhou},
  \citenamefont {Szczech}, \citenamefont {Pettes}, \citenamefont {Moore},
  \citenamefont {Jin},\ and\ \citenamefont {Shi}}]{phonon3}%
  \BibitemOpen
  \bibfield  {author} {\bibinfo {author} {\bibfnamefont {F.}~\bibnamefont
  {Zhou}}, \bibinfo {author} {\bibfnamefont {J.}~\bibnamefont {Szczech}},
  \bibinfo {author} {\bibfnamefont {M.~T.}\ \bibnamefont {Pettes}}, \bibinfo
  {author} {\bibfnamefont {A.~L.}\ \bibnamefont {Moore}}, \bibinfo {author}
  {\bibfnamefont {S.}~\bibnamefont {Jin}}, \ and\ \bibinfo {author}
  {\bibfnamefont {L.}~\bibnamefont {Shi}},\ }\href
  {http://dx.doi.org/10.1021/nl0706143} {\bibfield  {journal} {\bibinfo
  {journal} {Nano Letters}\ }\textbf {\bibinfo {volume} {7}},\ \bibinfo {pages}
  {1649} (\bibinfo {year} {2007})},\ \bibinfo {note} {pMID:
  17508772}\BibitemShut {NoStop}%
\bibitem [{\citenamefont {Zhou}\ \emph {et~al.}(2010)\citenamefont {Zhou},
  \citenamefont {Moore}, \citenamefont {Pettes}, \citenamefont {Lee},
  \citenamefont {Seol}, \citenamefont {Ye}, \citenamefont {Rabenberg},\ and\
  \citenamefont {Shi}}]{phonon4}%
  \BibitemOpen
  \bibfield  {author} {\bibinfo {author} {\bibfnamefont {F.}~\bibnamefont
  {Zhou}}, \bibinfo {author} {\bibfnamefont {A.~L.}\ \bibnamefont {Moore}},
  \bibinfo {author} {\bibfnamefont {M.~T.}\ \bibnamefont {Pettes}}, \bibinfo
  {author} {\bibfnamefont {Y.}~\bibnamefont {Lee}}, \bibinfo {author}
  {\bibfnamefont {J.~H.}\ \bibnamefont {Seol}}, \bibinfo {author}
  {\bibfnamefont {Q.~L.}\ \bibnamefont {Ye}}, \bibinfo {author} {\bibfnamefont
  {L.}~\bibnamefont {Rabenberg}}, \ and\ \bibinfo {author} {\bibfnamefont
  {L.}~\bibnamefont {Shi}},\ }\href@noop {} {\bibfield  {journal} {\bibinfo
  {journal} {Journal of Physics D: Applied Physics}\ }\textbf {\bibinfo
  {volume} {43}},\ \bibinfo {pages} {025406} (\bibinfo {year}
  {2010})}\BibitemShut {NoStop}%
\bibitem [{\citenamefont {Boukai}\ \emph {et~al.}(2008)\citenamefont {Boukai},
  \citenamefont {Bunimovich}, \citenamefont {Tahir-Kheli}, \citenamefont {Yu},
  \citenamefont {Goddard},\ and\ \citenamefont {Heath}}]{phonon5}%
  \BibitemOpen
  \bibfield  {author} {\bibinfo {author} {\bibfnamefont {A.~I.}\ \bibnamefont
  {Boukai}}, \bibinfo {author} {\bibfnamefont {Y.}~\bibnamefont {Bunimovich}},
  \bibinfo {author} {\bibfnamefont {J.}~\bibnamefont {Tahir-Kheli}}, \bibinfo
  {author} {\bibfnamefont {J.-K.}\ \bibnamefont {Yu}}, \bibinfo {author}
  {\bibfnamefont {W.~A.}\ \bibnamefont {Goddard}}, \ and\ \bibinfo {author}
  {\bibfnamefont {J.~R.}\ \bibnamefont {Heath}},\ }\href@noop {} {\bibfield
  {journal} {\bibinfo  {journal} {Nature}\ }\textbf {\bibinfo {volume} {451}},\
  \bibinfo {pages} {168} (\bibinfo {year} {2008})}\BibitemShut {NoStop}%
\bibitem [{\citenamefont {Chen}(1998)}]{chen3}%
  \BibitemOpen
  \bibfield  {author} {\bibinfo {author} {\bibfnamefont {G.}~\bibnamefont
  {Chen}},\ }\href {\doibase 10.1103/PhysRevB.57.14958} {\bibfield  {journal}
  {\bibinfo  {journal} {Phys. Rev. B}\ }\textbf {\bibinfo {volume} {57}},\
  \bibinfo {pages} {14958} (\bibinfo {year} {1998})}\BibitemShut {NoStop}%
\bibitem [{\citenamefont {Jordan}\ \emph
  {et~al.}(2013{\natexlab{a}})\citenamefont {Jordan}, \citenamefont {Sothmann},
  \citenamefont {S\'anchez},\ and\ \citenamefont {B\"uttiker}}]{sothmann1}%
  \BibitemOpen
  \bibfield  {author} {\bibinfo {author} {\bibfnamefont {A.~N.}\ \bibnamefont
  {Jordan}}, \bibinfo {author} {\bibfnamefont {B.}~\bibnamefont {Sothmann}},
  \bibinfo {author} {\bibfnamefont {R.}~\bibnamefont {S\'anchez}}, \ and\
  \bibinfo {author} {\bibfnamefont {M.}~\bibnamefont {B\"uttiker}},\ }\href
  {\doibase 10.1103/PhysRevB.87.075312} {\bibfield  {journal} {\bibinfo
  {journal} {Phys. Rev. B}\ }\textbf {\bibinfo {volume} {87}},\ \bibinfo
  {pages} {075312} (\bibinfo {year} {2013}{\natexlab{a}})}\BibitemShut
  {NoStop}%
\bibitem [{\citenamefont {Sothmann}\ \emph {et~al.}(2013)\citenamefont
  {Sothmann}, \citenamefont {Sánchez}, \citenamefont {Jordan},\ and\
  \citenamefont {Büttiker}}]{sothmann2}%
  \BibitemOpen
  \bibfield  {author} {\bibinfo {author} {\bibfnamefont {B.}~\bibnamefont
  {Sothmann}}, \bibinfo {author} {\bibfnamefont {R.}~\bibnamefont {Sánchez}},
  \bibinfo {author} {\bibfnamefont {A.~N.}\ \bibnamefont {Jordan}}, \ and\
  \bibinfo {author} {\bibfnamefont {M.}~\bibnamefont {Büttiker}},\ }\href
  {http://stacks.iop.org/1367-2630/15/i=9/a=095021} {\bibfield  {journal}
  {\bibinfo  {journal} {New Journal of Physics}\ }\textbf {\bibinfo {volume}
  {15}},\ \bibinfo {pages} {095021} (\bibinfo {year} {2013})}\BibitemShut
  {NoStop}%
\bibitem [{\citenamefont {Li}\ and\ \citenamefont
  {Jiang}(2016)}]{staircase_qdot}%
  \BibitemOpen
  \bibfield  {author} {\bibinfo {author} {\bibfnamefont {L.}~\bibnamefont
  {Li}}\ and\ \bibinfo {author} {\bibfnamefont {J.~H.}\ \bibnamefont {Jiang}},\
  }\href {http://dx.doi.org/10.1038/srep31974} {\bibfield  {journal} {\bibinfo
  {journal} {Scientific Reports}\ }\textbf {\bibinfo {volume} {6}} (\bibinfo
  {year} {2016})}\BibitemShut {NoStop}%
\bibitem [{\citenamefont {Singha}(2018)}]{aniket1}%
  \BibitemOpen
  \bibfield  {author} {\bibinfo {author} {\bibfnamefont {A.}~\bibnamefont
  {Singha}},\ }\href {\doibase https://doi.org/10.1016/j.physleta.2018.07.017}
  {\bibfield  {journal} {\bibinfo  {journal} {Physics Letters A}\ }\textbf
  {\bibinfo {volume} {382}},\ \bibinfo {pages} {3026 } (\bibinfo {year}
  {2018})}\BibitemShut {NoStop}%
\bibitem [{\citenamefont {Singha}\ \emph {et~al.}(2015)\citenamefont {Singha},
  \citenamefont {Mahanti},\ and\ \citenamefont {Muralidharan}}]{aniket}%
  \BibitemOpen
  \bibfield  {author} {\bibinfo {author} {\bibfnamefont {A.}~\bibnamefont
  {Singha}}, \bibinfo {author} {\bibfnamefont {S.~D.}\ \bibnamefont {Mahanti}},
  \ and\ \bibinfo {author} {\bibfnamefont {B.}~\bibnamefont {Muralidharan}},\
  }\href
  {http://scitation.aip.org/content/aip/journal/adva/5/10/10.1063/1.4933125}
  {\bibfield  {journal} {\bibinfo  {journal} {AIP Advances}\ }\textbf {\bibinfo
  {volume} {5}},\ \bibinfo {eid} {107210} (\bibinfo {year} {2015})}\BibitemShut
  {NoStop}%
\bibitem [{\citenamefont {Singha}\ and\ \citenamefont
  {Muralidharan}(2018)}]{aniket2}%
  \BibitemOpen
  \bibfield  {author} {\bibinfo {author} {\bibfnamefont {A.}~\bibnamefont
  {Singha}}\ and\ \bibinfo {author} {\bibfnamefont {B.}~\bibnamefont
  {Muralidharan}},\ }\href {\doibase 10.1063/1.5044254} {\bibfield  {journal}
  {\bibinfo  {journal} {Journal of Applied Physics}\ }\textbf {\bibinfo
  {volume} {124}},\ \bibinfo {pages} {144901} (\bibinfo {year}
  {2018})}\BibitemShut {NoStop}%
\bibitem [{\citenamefont {Singha}\ and\ \citenamefont
  {Muralidharan}(2017)}]{aniket3}%
  \BibitemOpen
  \bibfield  {author} {\bibinfo {author} {\bibfnamefont {A.}~\bibnamefont
  {Singha}}\ and\ \bibinfo {author} {\bibfnamefont {B.}~\bibnamefont
  {Muralidharan}},\ }\href {\doibase 10.1038/s41598-017-07935-w} {\bibfield
  {journal} {\bibinfo  {journal} {Scientific Reports}\ }\textbf {\bibinfo
  {volume} {7}},\ \bibinfo {pages} {7879} (\bibinfo {year} {2017})}\BibitemShut
  {NoStop}%
\bibitem [{\citenamefont {Humphrey}\ and\ \citenamefont
  {Linke}(2005)}]{Humphrey}%
  \BibitemOpen
  \bibfield  {author} {\bibinfo {author} {\bibfnamefont {T.~E.}\ \bibnamefont
  {Humphrey}}\ and\ \bibinfo {author} {\bibfnamefont {H.}~\bibnamefont
  {Linke}},\ }\href {\doibase 10.1103/PhysRevLett.94.096601} {\bibfield
  {journal} {\bibinfo  {journal} {Phys. Rev. Lett.}\ }\textbf {\bibinfo
  {volume} {94}},\ \bibinfo {pages} {096601} (\bibinfo {year}
  {2005})}\BibitemShut {NoStop}%
\bibitem [{\citenamefont {Humphrey}\ \emph {et~al.}(2002)\citenamefont
  {Humphrey}, \citenamefont {Newbury}, \citenamefont {Taylor},\ and\
  \citenamefont {Linke}}]{Humphrey2}%
  \BibitemOpen
  \bibfield  {author} {\bibinfo {author} {\bibfnamefont {T.~E.}\ \bibnamefont
  {Humphrey}}, \bibinfo {author} {\bibfnamefont {R.}~\bibnamefont {Newbury}},
  \bibinfo {author} {\bibfnamefont {R.~P.}\ \bibnamefont {Taylor}}, \ and\
  \bibinfo {author} {\bibfnamefont {H.}~\bibnamefont {Linke}},\ }\href
  {\doibase 10.1103/PhysRevLett.89.116801} {\bibfield  {journal} {\bibinfo
  {journal} {Phys. Rev. Lett.}\ }\textbf {\bibinfo {volume} {89}},\ \bibinfo
  {pages} {116801} (\bibinfo {year} {2002})}\BibitemShut {NoStop}%
\bibitem [{\citenamefont {Yamamoto}\ \emph {et~al.}(2016)\citenamefont
  {Yamamoto}, \citenamefont {Entin-Wohlman}, \citenamefont {Aharony},\ and\
  \citenamefont {Hatano}}]{yamamoto2}%
  \BibitemOpen
  \bibfield  {author} {\bibinfo {author} {\bibfnamefont {K.}~\bibnamefont
  {Yamamoto}}, \bibinfo {author} {\bibfnamefont {O.}~\bibnamefont
  {Entin-Wohlman}}, \bibinfo {author} {\bibfnamefont {A.}~\bibnamefont
  {Aharony}}, \ and\ \bibinfo {author} {\bibfnamefont {N.}~\bibnamefont
  {Hatano}},\ }\href {\doibase 10.1103/PhysRevB.94.121402} {\bibfield
  {journal} {\bibinfo  {journal} {Phys. Rev. B}\ }\textbf {\bibinfo {volume}
  {94}},\ \bibinfo {pages} {121402} (\bibinfo {year} {2016})}\BibitemShut
  {NoStop}%
\bibitem [{\citenamefont {Yamamoto}\ and\ \citenamefont
  {Hatano}(2015)}]{yamamoto1}%
  \BibitemOpen
  \bibfield  {author} {\bibinfo {author} {\bibfnamefont {K.}~\bibnamefont
  {Yamamoto}}\ and\ \bibinfo {author} {\bibfnamefont {N.}~\bibnamefont
  {Hatano}},\ }\href {\doibase 10.1103/PhysRevE.92.042165} {\bibfield
  {journal} {\bibinfo  {journal} {Phys. Rev. E}\ }\textbf {\bibinfo {volume}
  {92}},\ \bibinfo {pages} {042165} (\bibinfo {year} {2015})}\BibitemShut
  {NoStop}%
\bibitem [{\citenamefont {Kim}\ and\ \citenamefont {Lundstrom}(2012)}]{kim1}%
  \BibitemOpen
  \bibfield  {author} {\bibinfo {author} {\bibfnamefont {R.}~\bibnamefont
  {Kim}}\ and\ \bibinfo {author} {\bibfnamefont {M.~S.}\ \bibnamefont
  {Lundstrom}},\ }\href
  {http://scitation.aip.org/content/aip/journal/jap/111/2/10.1063/1.3678001}
  {\bibfield  {journal} {\bibinfo  {journal} {Journal of Applied Physics}\
  }\textbf {\bibinfo {volume} {111}},\ \bibinfo {eid} {024508} (\bibinfo {year}
  {2012})}\BibitemShut {NoStop}%
\bibitem [{\citenamefont {Kim}\ and\ \citenamefont {Lundstrom}(2011)}]{kim2}%
  \BibitemOpen
  \bibfield  {author} {\bibinfo {author} {\bibfnamefont {R.}~\bibnamefont
  {Kim}}\ and\ \bibinfo {author} {\bibfnamefont {M.~S.}\ \bibnamefont
  {Lundstrom}},\ }\href
  {http://scitation.aip.org/content/aip/journal/jap/110/3/10.1063/1.3619855}
  {\bibfield  {journal} {\bibinfo  {journal} {Journal of Applied Physics}\
  }\textbf {\bibinfo {volume} {110}},\ \bibinfo {eid} {034511} (\bibinfo {year}
  {2011})}\BibitemShut {NoStop}%
\bibitem [{\citenamefont {Thesberg}\ \emph {et~al.}(2016)\citenamefont
  {Thesberg}, \citenamefont {Kosina},\ and\ \citenamefont {Neophytou}}]{neo}%
  \BibitemOpen
  \bibfield  {author} {\bibinfo {author} {\bibfnamefont {M.}~\bibnamefont
  {Thesberg}}, \bibinfo {author} {\bibfnamefont {H.}~\bibnamefont {Kosina}}, \
  and\ \bibinfo {author} {\bibfnamefont {N.}~\bibnamefont {Neophytou}},\ }\href
  {\doibase 10.1063/1.4972192} {\bibfield  {journal} {\bibinfo  {journal}
  {Journal of Applied Physics}\ }\textbf {\bibinfo {volume} {120}},\ \bibinfo
  {pages} {234302} (\bibinfo {year} {2016})}\BibitemShut {NoStop}%
\bibitem [{\citenamefont {Nakpathomkun}\ \emph {et~al.}(2010)\citenamefont
  {Nakpathomkun}, \citenamefont {Xu},\ and\ \citenamefont
  {Linke}}]{nakpathomkun}%
  \BibitemOpen
  \bibfield  {author} {\bibinfo {author} {\bibfnamefont {N.}~\bibnamefont
  {Nakpathomkun}}, \bibinfo {author} {\bibfnamefont {H.~Q.}\ \bibnamefont
  {Xu}}, \ and\ \bibinfo {author} {\bibfnamefont {H.}~\bibnamefont {Linke}},\
  }\href {http://link.aps.org/doi/10.1103/PhysRevB.82.235428} {\bibfield
  {journal} {\bibinfo  {journal} {Phys. Rev. B}\ }\textbf {\bibinfo {volume}
  {82}},\ \bibinfo {pages} {235428} (\bibinfo {year} {2010})}\BibitemShut
  {NoStop}%
\bibitem [{\citenamefont {Jordan}\ \emph
  {et~al.}(2013{\natexlab{b}})\citenamefont {Jordan}, \citenamefont {Sothmann},
  \citenamefont {S\'anchez},\ and\ \citenamefont {B\"uttiker}}]{jordan1}%
  \BibitemOpen
  \bibfield  {author} {\bibinfo {author} {\bibfnamefont {A.~N.}\ \bibnamefont
  {Jordan}}, \bibinfo {author} {\bibfnamefont {B.}~\bibnamefont {Sothmann}},
  \bibinfo {author} {\bibfnamefont {R.}~\bibnamefont {S\'anchez}}, \ and\
  \bibinfo {author} {\bibfnamefont {M.}~\bibnamefont {B\"uttiker}},\
  }\href@noop {} {\bibfield  {journal} {\bibinfo  {journal} {Phys. Rev. B}\
  }\textbf {\bibinfo {volume} {87}},\ \bibinfo {pages} {075312} (\bibinfo
  {year} {2013}{\natexlab{b}})}\BibitemShut {NoStop}%
\bibitem [{\citenamefont {Choi}\ and\ \citenamefont {Jordan}(2016)}]{jordan2}%
  \BibitemOpen
  \bibfield  {author} {\bibinfo {author} {\bibfnamefont {Y.}~\bibnamefont
  {Choi}}\ and\ \bibinfo {author} {\bibfnamefont {A.~N.}\ \bibnamefont
  {Jordan}},\ }\href@noop {} {\bibfield  {journal} {\bibinfo  {journal}
  {Physica E}\ }\textbf {\bibinfo {volume} {74}},\ \bibinfo {pages} {465}
  (\bibinfo {year} {2016})}\BibitemShut {NoStop}%
\bibitem [{\citenamefont {Sothmann}\ \emph {et~al.}(2014)\citenamefont
  {Sothmann}, \citenamefont {S\'{a}nchez},\ and\ \citenamefont
  {Jordan}}]{sothmann}%
  \BibitemOpen
  \bibfield  {author} {\bibinfo {author} {\bibfnamefont {B.}~\bibnamefont
  {Sothmann}}, \bibinfo {author} {\bibfnamefont {R.}~\bibnamefont
  {S\'{a}nchez}}, \ and\ \bibinfo {author} {\bibfnamefont {A.~N.}\ \bibnamefont
  {Jordan}},\ }\href@noop {} {\bibfield  {journal} {\bibinfo  {journal}
  {Nanotechnology}\ }\textbf {\bibinfo {volume} {26}},\ \bibinfo {pages}
  {32001} (\bibinfo {year} {2014})}\BibitemShut {NoStop}%
\bibitem [{\citenamefont {Muralidharan}\ and\ \citenamefont
  {Grifoni}(2012)}]{bm}%
  \BibitemOpen
  \bibfield  {author} {\bibinfo {author} {\bibfnamefont {B.}~\bibnamefont
  {Muralidharan}}\ and\ \bibinfo {author} {\bibfnamefont {M.}~\bibnamefont
  {Grifoni}},\ }\href@noop {} {\bibfield  {journal} {\bibinfo  {journal} {Phys.
  Rev. B}\ }\textbf {\bibinfo {volume} {85}},\ \bibinfo {pages} {155423}
  (\bibinfo {year} {2012})}\BibitemShut {NoStop}%
\bibitem [{\citenamefont {Agarwal}\ and\ \citenamefont
  {Muralidharan}(2014)}]{agarwal}%
  \BibitemOpen
  \bibfield  {author} {\bibinfo {author} {\bibfnamefont {A.}~\bibnamefont
  {Agarwal}}\ and\ \bibinfo {author} {\bibfnamefont {B.}~\bibnamefont
  {Muralidharan}},\ }\href@noop {} {\bibfield  {journal} {\bibinfo  {journal}
  {App. Phys. Lett.}\ }\textbf {\bibinfo {volume} {105}},\ \bibinfo {pages}
  {013104} (\bibinfo {year} {2014})}\BibitemShut {NoStop}%
\bibitem [{\citenamefont {De}\ and\ \citenamefont {Muralidharan}(2016)}]{BD}%
  \BibitemOpen
  \bibfield  {author} {\bibinfo {author} {\bibfnamefont {B.}~\bibnamefont
  {De}}\ and\ \bibinfo {author} {\bibfnamefont {B.}~\bibnamefont
  {Muralidharan}},\ }\href {\doibase 10.1103/PhysRevB.94.165416} {\bibfield
  {journal} {\bibinfo  {journal} {Phys. Rev. B}\ }\textbf {\bibinfo {volume}
  {94}},\ \bibinfo {pages} {165416} (\bibinfo {year} {2016})}\BibitemShut
  {NoStop}%
\bibitem [{\citenamefont {Zimbovskaya}(2016)}]{zimb}%
  \BibitemOpen
  \bibfield  {author} {\bibinfo {author} {\bibfnamefont {N.~A.}\ \bibnamefont
  {Zimbovskaya}},\ }\href@noop {} {\bibfield  {journal} {\bibinfo  {journal}
  {Journal of Physics: Condensed Matter}\ }\textbf {\bibinfo {volume} {28}},\
  \bibinfo {pages} {183002} (\bibinfo {year} {2016})}\BibitemShut {NoStop}%
\bibitem [{\citenamefont {Leijnse}\ \emph {et~al.}(2010)\citenamefont
  {Leijnse}, \citenamefont {Wegewijs},\ and\ \citenamefont {Flensberg}}]{leij}%
  \BibitemOpen
  \bibfield  {author} {\bibinfo {author} {\bibfnamefont {M.}~\bibnamefont
  {Leijnse}}, \bibinfo {author} {\bibfnamefont {M.~R.}\ \bibnamefont
  {Wegewijs}}, \ and\ \bibinfo {author} {\bibfnamefont {K.}~\bibnamefont
  {Flensberg}},\ }\href@noop {} {\bibfield  {journal} {\bibinfo  {journal}
  {Phys. Rev. B}\ }\textbf {\bibinfo {volume} {82}},\ \bibinfo {pages} {045412}
  (\bibinfo {year} {2010})}\BibitemShut {NoStop}%
\end{thebibliography}%
\end{document}